\documentclass[twocolumn,showpacs,preprintnumbers,amsmath,amssymb]{revtex4}
\usepackage{graphicx}% Include figure files
\usepackage{dcolumn}% Align table columns on decimal point
\usepackage{bm}% bold math
\usepackage{float}

\usepackage{color}
\usepackage{refstyle}
\usepackage{mathrsfs}
\usepackage{esint}
\usepackage[unicode=true,pdfusetitle,bookmarks=true,
bookmarksnumbered=false,bookmarksopen=false,breaklinks=false,
pdfborder={0 0 0},backref=false,colorlinks=true,citecolor=red]
{hyperref}
\usepackage{cleveref}
\providecommand\eqref[1]{\ref{eq:#1}}

\renewcommand\epsilon{\varepsilon}

\allowdisplaybreaks[3]
\newcommand\mcm[1]{{{#1}}}

\begin{document}

\title{Probing the shear viscosity of an active nematic}% Force line breaks with \\

\author{Pau Guillamat$^a$}
\author{Jordi Ign\'es-Mullol$^a$}
\author{Suraj Shankar$^b$}
\author{M. Cristina Marchetti$^b$}
\author{Francesc Sagu\'es$^a$}
\email{f.sagues@ub.edu}
\affiliation{$^a$Departament de Qu\'{\i}mica F\'{\i}sica and Institute of Nanoscience and Nanotechnology (IN2UB), Universitat de Barcelona, Mart\'{\i} i Franqu\`es 1, 08028 Barcelona, Catalonia, Spain.\\
$^b$Physics Department and Syracuse Soft Matter Program, Syracuse University, Syracuse, NY 13244, USA.}

\date{\today}% It is always \today, today,
             %  but any date may be explicitly specified

\begin{abstract}
	\emph{In vitro }reconstituted active systems, such as the ATP-driven microtubule bundle suspension developed by the Dogic group, provide a fertile testing ground for elucidating the phenomenology of active liquid crystalline states. Controlling such novel phases of matter crucially depends on our knowledge of their material and physical properties. In this letter, we show that  the rheological properties of an active nematic film can be probed by varying its hydrodynamic coupling to a bounding oil layer. Using the motion of disclinations as intrinsic tracers of the flow field and a hydrodynamic model, we obtain \mcm{an estimate} for the as of now unknown shear viscosity of the nematic film. Knowing this now provides us with an additional handle for robust and precision tunable control of the emergent dynamics of active fluids.
\end{abstract}

\pacs{87.16.Ka,61.30.Jf,83.85.Jn}

\maketitle

%\section{Introduction}
%\label{sec:introduction}

Active systems are collections of self-propelling entities that display non-equilibrium self-organization  on many scales~\cite{Marchetti13,Ramaswamy10}.
%Soft active matter has emerged during these last years as a new paradigm in the field of non-equilibrium Condensed Matter Physics \cite{Marchetti13, Ramaswamy10}.
Examples from the living world include animal flocks \cite{Cavagna10}, bacterial colonies \cite {Dombrowski04, Zhang10}, living tissues and cytoskeletal extracts~\cite{Schaller10,Sumino12,Sanchez12}. Ingenious synthetic analogues composed either of externally driven \cite{Narayan07,Bricard13, Hernandez14} or autonomously propelled elements \cite{Paxton04,Howse07, Palacci13} have also been developed. The distinctive feature that unifies these systems is that they are composed of interacting units that convert ambient or stored energy into self-sustained motion, from coherently organized to seemingly chaotic.

Reconstituted suspensions of cytoskeletal filaments and associated motor proteins are active systems that have proven ideal for quantitative studies of the origin of subcellular organization, from the contractility of the actomyosin cytoskeleton to the division of the mitotic spindle. The group of Z. Dogic \cite{Sanchez12,Henkin14} has pioneered a remarkable model system of an active gel consisting of a suspension of microtubules (MTs) in the presence of Adenosin Triphosphate (ATP)-fuelled kinesin motors. Bundled microtubules behave as active units that exert extensile forces on their environment and are capable of reproducing \emph{in vitro} some of the unique behavior of living systems. When concentrated at an oil-water interface, the suspension of MT bundles organizes into an active nematic that exhibits self-sustained spontaneous flows with striking resemblance to the streaming used by cells to circulate their fluid content. At high enough activity active turbulent flows develop, with proliferation of unbound disclinations - the distinctive textures of two-dimensional films of nematic liquid crystals. By confining the suspension of MTs to the surface of a lipid vesicle, Keber \emph{et al.} fabricated ``active vesicles" that can undergo spontaneous oscillations and remarkable shape changes~\cite{Keber14}. Finally, droplets of active gels squeezed in a narrow gap between two glass plates perform erratic autonomous motion \cite{Sanchez12}.

While a lot of studies have focused on modeling this rich living-like behavior and especially on the generation and dynamics of topological defects, the material properties of active nematics remain largely unexplored. One exception is a very recent work by some of us, where an anisotropic shear viscosity, arising from a contacting passive thermotropic liquid crystal, is used to align active nematic flows \cite{Guillamat16}.

Here, we examine quantitatively the influence of the combined hydrodynamics of the $2d$ active nematic and the bounding bulk passive fluids on the active dynamics, and use it to infer an estimate for the shear viscosity of the active material. This is obtained through a simple hydrodynamic model that examines the flow induced by nematic textures confined at the interface of two fluids of different viscosity \cite{Saffman75,Saffman76,Lubensky96}. A fit of the model's prediction for the defect velocity when the oil viscosity is varied over five orders of magnitude allows us to \mcm{determine an effective bulk shear viscosity of the active nematic and to infer the viscosity of the film.} We show that the rheological characteristics of the interface also have a profound effect on the textures and flows of the active nematic.
% In addition to evidence a marked effect of the rheological characteristics of the interface on the textures and flows of the bundled MTs, a hydrodynamic model will permit to obtain, for the first time, a value for the two-dimensional viscosity of this active material.
%
\begin{figure}
  \includegraphics[width=0.9\columnwidth]{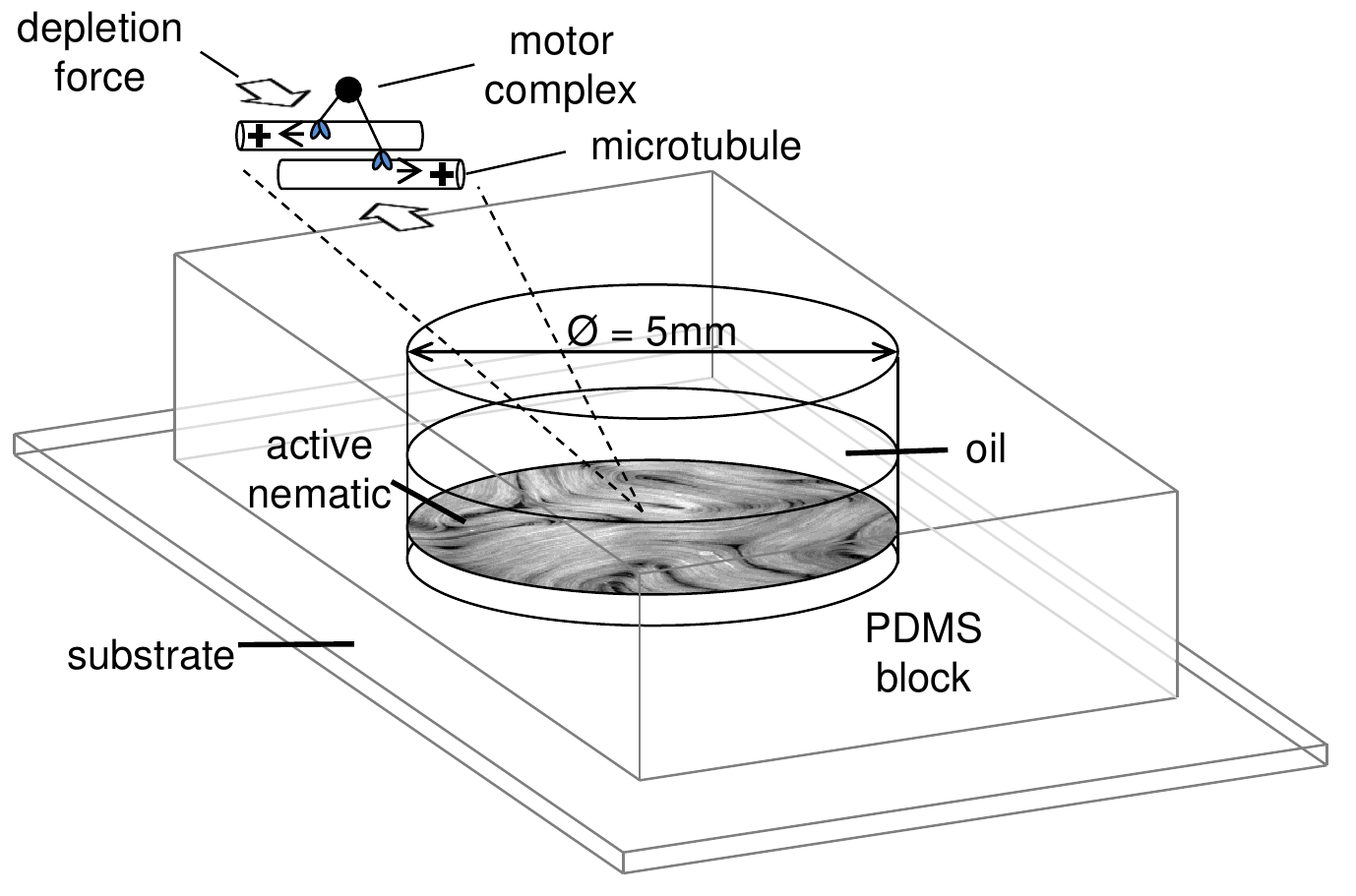}
  \caption{ Experimental setup. The liquids are contained in a cylindrical well custom-engineered on a PDMS block that is bound to a substrate coated with a polymer brush that prevents protein adhesion. The aqueous active material is injected below a volume of silicone oil. Depletion forces promote the formation of the MT bundles, sheared by motor clusters, and lead to the condensation of the active nematic film at the oil/water interface.}
  \label{fig:cell}
\end{figure}

The active material we study is based on the hierarchical self-assembly of tubulin into stabilized micron-length fluorescent MTs, organized into bundles that are internally cross-linked and sheared by clusters of kinesin motors ~\cite{Sanchez12}. This leads to MT bundle elongation, bending and buckling, which results in extensile local stresses on the surrounding fluid. Once depleted towards a surfactant-decorated oil/water interface, the kinesin/tubulin gel develops the well-known active nematic configuration, which is characterized by self-sustained flows and orientational order of the aligned filaments. Although the thickness of the active nematic layer is not known with precision, we estimate it to be in the range between the minimal bundle thickness, $0.2\;\mu{\rm m}$, estimated from the typical sizes of filaments and motor proteins, and the resolution of fluorescence confocal micrographs, $2\;\mu{\rm m}$. Our open-cell arrangement is based on a custom polydimethylsiloxane (PDMS) block containing a cylindrical well, and bound to a support plate (see Fig.~\ref{fig:cell} and \cite{Guillamat16}). After filling the well with silicone oil of the desired viscosity, the aqueous gel is injected between the bottom plate and the oil. This results in an aqueous layer of $100-200\;\mu{\rm m}$ depth underneath an oil phase of $1-2\;{\rm mm}$ depth. Unlike the original arrangement~\cite{Sanchez12}, our setup does not demand the use of a low viscosity oil, thus allowing us to explore nearly five orders of magnitude of viscosity contrast between the interfacing oil and the aqueous bulk. In the experiments reported here, ATP concentration is kept at 1.4 mM by means of an enzymatic ATP-regenerator that is incorporated in the active material, which sets a constant activity for the active material over the time scale of the experiment.

\begin{figure}
  % Requires \usepackage{graphicx}
  \includegraphics[width=8cm]{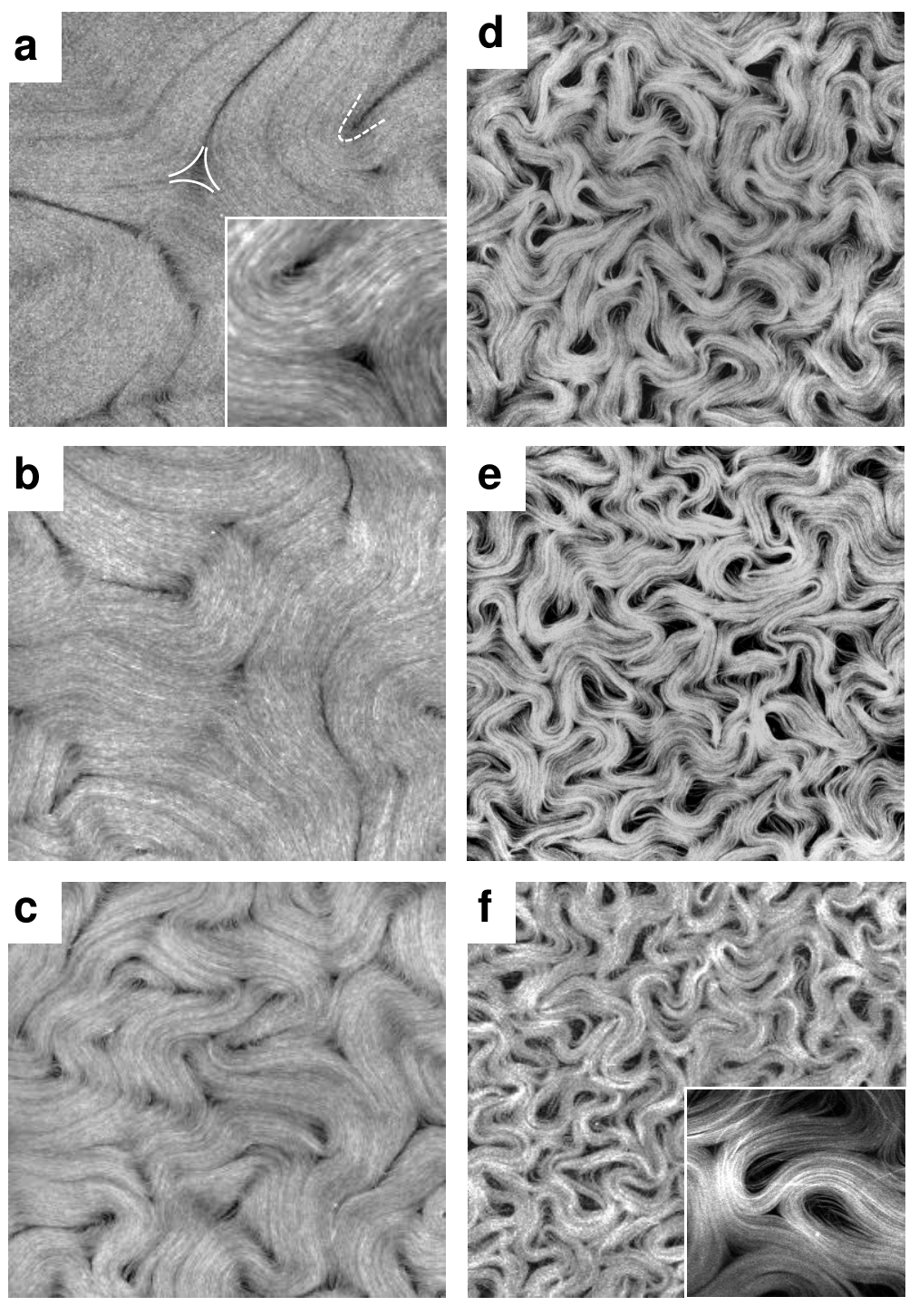}\\
  \caption{Active nematic in contact with silicone oils of different viscosity. Fluorescence confocal micrographs are 400$\mu$m wide. The oil viscosity, from (a) to (f), is $5\times 10^{-3}$, $5\times 10^{-2}$, $0.5$, $5$, $12.5$, and $300$ Pa s, respectively. In frame (a) two streamlines for defects with parabolic ($+1/2$, dashed) and hyperbolic ($-1/2$, solid) morphologies are highlighted. Insets show sections at four-fold magnification.}
  \label{fig_experiments}
\end{figure}

\emph{Results.} The spontaneous flow is the result of the extension of the elongated fibers from continuous microtubule sliding driven by the kinesin motors. Because of their extensile nature, MT bundles bend and form parabolic folds, bounding dark regions devoid of MTs (Fig. \ref{fig_experiments}.a). The orientation of the aligned filaments performs half a turn along any closed circuit surrounding either the tip or the tail of the fold. This result allows us to associate a defect topological charge $+1/2$ to the parabolic tip of the fold, and a charge $-1/2$ to the hyperbolic tail. As topological defects are created and annihilated in pairs, the total topological charge adds up to zero at all times. Since the flow originates at the tip of the folds, $+1/2$ defects become like active or self-propelled particles, as first quantified in~\cite{Giomi13}, and can be used as intrinsic tracers for the active flow. On the other hand, $-1/2$ defects often occupy flow stagnation points (Fig. 2 in SI). In these nematic films the $+1/2$ defects have also been observed to exhibit long-range nematic alignment~ \cite{decamp15}, although the mechanism for such alignment is still controversial \cite{decamp15,DunkelOza,Yeomans16}.

Our study has revealed a clear influence of the oil viscosity, $\eta_o$, on the morphology and dynamics of the active nematic. In Fig. \ref{fig_experiments}, we show snapshots of the active nematic in contact with oils of different viscosity, for the same activity (i.e., concentration of ATP). These patterns are characterized by the proliferation of randomly moving defects, organizing into the so-called active turbulent regime \cite{Giomi13, Shi13, Weber14, Thampi14, Giomi14, Giomi15}. It is clear that at higher $\eta_o$ the number of defects increases and degrades the orientational order of the filament bundles. At the same time, the speed of defects decreases for increasing $\eta_o$ (Video S1).
It is also apparent that textures in contact with oils of smaller viscosity appear less fluorescent (Fig. 3 in SI) and more tenuous, as compared to those observed for higher $\eta_o$. This indicates that a large $\eta_o$ concentrates the MT bundles while also amplifying the typical size of the empty regions that are the cores of the defect textures.
Additionally, we observe an  increase of the radius of curvature of microtubule bending with $\eta_o$, as evidenced when comparing the insets of Figs \ref{fig_experiments}a and \ref{fig_experiments}f. Since the defect core length, denoted by $\xi_Q$, is expected to grow as $\sqrt{K}$, with $K$ the nematic bending rigidity, based on these observations we conjecture that the nematic material becomes stiffer in contact with oils of increasing viscosity.
Textures such as the one displayed in Fig.~\ref{fig_experiments}f are rather disordered in the sense that the nematic order parameter would average to a small value. The system, however, still shows a characteristic structure that can be analyzed in terms of the topological defects of a nematic liquid crystal.

\begin{figure}
  % Requires \usepackage{graphicx}
  \includegraphics[width=0.9\columnwidth]{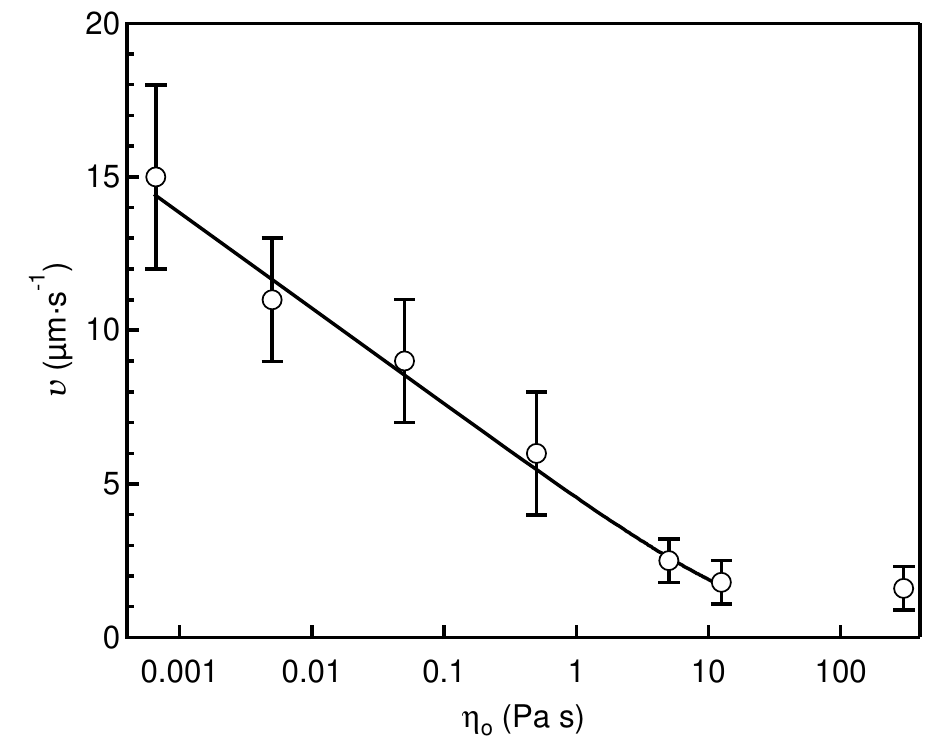}
  \caption{Average speed of $+1/2$ defects as a function of oil viscosity. The line is a fit according to the hydrodynamic model described in the text.}
  \label{fig_velocity}
\end{figure}

\begin{figure}
\includegraphics[width=0.9\columnwidth]{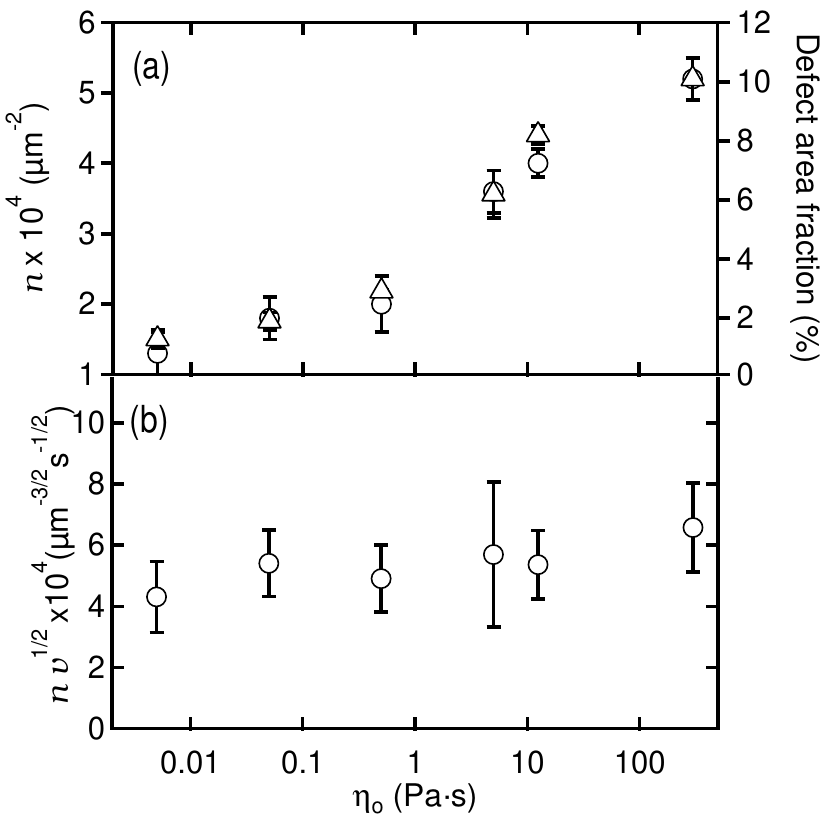}
\caption{(a) Defect density ($\bigcirc$) and area fraction covered by defects ($\triangle$) as a function of oil viscosity.(b) The combination $n v^{1/2}$ is essentially independent of oil viscosity, supporting the scaling ansatz $n\sim v^{-1/2}$ of defect density with speed discussed in the text.
}
 \label{fig_defect_density}
\end{figure}

We have quantified both the number density, $n$, and the velocity, $v$, of the  $+1/2$ defects for the realizations shown in Fig. \ref{fig_experiments}. We observe that $v$ decreases logarithmically with $\eta_o$ before a saturation is reached for the most viscous oils (Fig. \ref{fig_velocity}). This data is successfully fitted with the hydrodynamic model described below, which allows us to estimate the viscosity of the active nematic. Conversely, the defect density grows steadily with $\eta_o$ (Fig. \ref{fig_defect_density}a). We have also measured the average size of the defect core, defined as the region devoid of MTs surrounding each defect. We observe that the area fraction occupied by the defect cores is proportional to the defect density, and the average defect core area, which we relate to $\xi_Q$, grows with $\eta_o$ until saturation for the most viscous oils (see Fig. 4 in SI).

The measurements of $v$ and $n$ can be combined to reveal a simple scaling relationship between these two quantities. The scaling ansatz can be obtained using generic arguments already proposed for freely suspended active nematics~\cite{Thampi13}. We assume a simple rate equation for the defect density, $dn/dt=R_c-R_a$, with $R_c$ and $R_a$ the rates of defect creation and annihilation per unit area. The creation rate can be estimated as $R_c\sim (\ell_{\alpha}^2\tau)^{-1}$, where $\ell_{\alpha}=\sqrt{K/|\alpha|}$ is the active length scale determined by balancing active and elastic stresses, with $\alpha$ an active stress scale controlled by ATP concentration and chosen negative for extensile systems. It has been shown that this length scale controls the spatial correlations in the active turbulent regimes of `wet' active nematics~\cite{Giomi15,Hemingway16}. The time scale $\tau={\gamma}/{|\alpha|}$ controls the relaxation of the director upon distortion by active stresses, with $\gamma$ the rotational viscosity of the nematic. The annihilation rate can be estimated as $R_a\sim \sigma v n^2$, with $\sigma$ an effective cross section (a length in  two-dimensions) that quantifies the range of defect interactions. At steady state, the average number density of defects remains constant, giving the scaling prediction $v n^2 \sim \alpha^2/(\gamma \sigma K)$. For a constant ATP concentration, $\alpha$ and $\gamma$ will be constant, leading to the simplified scaling $v n^2 \sim 1/(\sigma K)$.  Viscous stresses at the interface effectively dampen the active nematic speed (Fig.~\ref{fig_velocity}), which results in more sporadic defect annihilation events. Since the ATP concentration is kept constant in all these experiments, the defect creation rate is sustained. As a result, the number of defects increases (Fig.~\ref{fig_defect_density}a).
Notice that the above scaling implies that the product $\sigma K$ is independent of $\eta_o$ (Fig.~\ref{fig_defect_density}b).   Taking into account the experimental evidence that $K$ increases with $\eta_o$, this scaling behavior suggests that the range of defect interaction decreases in contact with more viscous oils. This trend is opposite to that exhibited by the defect core size, so clearly $\sigma$ and $\xi_Q$ must be independent length scales. As a matter of fact, we may quantify the change of $\sigma$ with $\eta_o$ by replacing $K$ by $\xi_Q^2$ in the scaling above, revealing that $\sigma$ decreases by a factor of four in the explored range of oil viscosities (Fig. 5 in SI).

%Creation rate results from combining typical area and time scales for defect formation. The former is assumed to be given by the active length scale $l_a=\sqrt{K/{\alpha}}$ \cite{Giomi15}, while the time unit is fixed by the director relaxation when distorted by active stresses $\tau= {\gamma}/{\alpha}$. Therefore, defect creation rate is given as $v_{cre}\simeq\frac{\alpha^2}{\gamma K}$, and annihilation, $v_{ann}\simeq  v n^2$. Equating both results leads to a simple scaling prediction for the number density of defects $n\simeq\frac{\alpha}{\sqrt{v}}$.
%In these expressions, $\alpha$ is the activity parameter, which we consider to be linearly related to  $\ln [ATP]$, $\gamma$ denotes the rotational viscosity coefficient, and $K$ stands for the elastic constant of the 2D nematic. The observed scaling indicates that $\alpha$ is constant in the full range of used oil viscosities.
%As shown in the hydrodynamic model,%

%\section{The hydrodynamic model}
%\label{sec:model}

\emph{Hydrodynamic model.} To capture the effect of viscous stresses propagated in the nematic by the viscosity contrast at the oil/water interface, we consider the hydrodynamics of a thin active nematic layer confined between two bulk fluids (oil and water) and calculate the velocity that the director distortion due to a $+1/2$ disclination creates at the core of the defect. Following Ref.~\cite{Giomi13}, and consistently with the analysis of the experiments, we identify this with the defect velocity. We find that the presence of the bulk fluids qualitatively changes the defect velocity as compared to \mcm{the previously considered cases of a free-standing nematic layer~\cite{Giomi13,Giomi14},  a finite-thickness layer of bulk nematic~\cite{Toth02,Thampi14}, and a nematic layer with frictional damping from a substrate~\cite{Pismen13}.}

To evaluate the defect velocity, we start with coupled Stokes equations for oil, water and nematic layer,
\begin{eqnarray}
\label{eq:Stokes-a}
&&\eta_{i}\nabla^2\mathbf{u}_i-\bm\nabla p_i=0\ ,\\
\label{eq:Stokes-N}
&&\eta_N\nabla^2\mathbf{u}_\perp-\bm\nabla_\perp p+\mathbf{\hat{n}}\cdot(\bm\sigma_o-\bm\sigma_w)+\bm\nabla_\perp\cdot\bm\sigma_a=0\;,
\end{eqnarray}
where $\eta_i$, $\mathbf{u}_i$ and $p_i$, with $i=o,w$ denoting oil or water, are the $3d$ viscosity, flow velocity, and pressure of the oil and water subphases respectively; $\bm\sigma_o-\bm\sigma_w$ is the stress jump across the interface, which is projected onto the unit interface normal $\mathbf{\hat{n}}$; $\eta_N$, $\mathbf{u}_\perp$ and $p$ are the $2d$ viscosity, in-plane flow velocity and lateral pressure of the active nematic layer. Finally, $\bm\sigma_a=\alpha\bm Q$ is the active stress arising from the extensile force dipoles exerted by MT bundles on their surroundings and proportional to the nematic alignment tensor $\bm Q$. We compute the nematic flow field $\mathbf{u}_\perp$ due to stationary textures of the order parameter $\bm Q$ corresponding to either a $+1/2$ or a $-1/2$ defect by solving~\eqref{Stokes-a,Stokes-N} for an incompressible flow with vanishing of the normal velocity and continuity  of the tangential velocity at the nematic interface, located at $z=0$.
%The program of finding the defect velocity is as follows. We begin with a stationary configuration of the nematic field (in particular, we restrict ourselves to single $\pm1/2$ disclinations in the plane). Using the nematic alignment tensor $Q$ corresponding to this configuration, the associated active stress is computed as $\sigma^a=\alpha\del\cdot Q$ where $\alpha$ is the phenomenological activity parameter. This stress provides the source for generating active backflow, which is then computed using the coupled multiphase Stokes equations.
%As the hydrodynamic part of the problem (i.e the coupled incompressible Stokes equations) is linear, it can be solved once and for all to give the hydrodynamic propagator (the green's function or non-local mobility tensor).
Thanks to the linearity of the Stokes equations, the solution is easily written in Fourier space in term of a Green's function.
As the depth of both the oil layer and the bulk fluid subphases are much larger than the thickness of the active nematic layer, we consider both bulk layers to be semi-infinite. In this limit, the Fourier components of the flow velocity in the nematic layer \mcm{are} $\mathbf{u}_\perp(\mathbf{k})=G(k)\mathcal{P}\mathbf{f}(\mathbf{k})$ , where $\mathcal{P}=\bm{I}-\mathbf{k}\mathbf{k}/k^2$ is a transverse projection operator, $\mathbf{f}(\mathbf{k})=\int_{\mathbf{r}}e^{-i\mathbf{k}\cdot\mathbf{r}}\bm\nabla_\perp\cdot\bm\sigma_a$ and the Green's function is given by
%
%\begin{equation}
	${G}(k)=\left[\eta_{N}k^2+\eta_ok\right]^{-1}$  \cite{Lubensky96}.
%\label{eq_greens}
%\end{equation}
%
\mcm{The} length scale $\ell_{\eta}=\eta_{N}/(\eta_o+\eta_w)\simeq \eta_{N}/\eta_o$ (for $\eta_w\ll\eta_o$)  \mcm{controls} the crossover from two dimensional surface flows to three dimensional bulk dominated flows. %, \mcm{and} we have neglected the viscosity of water \mcm{compared} to that of the oil.
%Fourier transforming Eq. \ref{eq_greens} into real space, we find, for $r\ll\ell_{\eta}$ (for large $r$ we will eventually be cutoff by the transverse confinement of the oil and water layers),
%\begin{align}
%%	G(\vec{r})&=\int\dd^2k\dfrac{e^{i\vec{k}\cdot\vec{r}}}{\eta_{AN}(k^2+ak)}\sim\dfrac{1}{2\pi\eta_{AN}}\int_{0}^{1/r}\dfrac{\dd k}{(k+a)}\nonumber\\
%%						  &=\dfrac{1}{2\pi\eta_{AN}}\log\left(\dfrac{1}{ar}\right)
%	G(r)=\int\frac{\dd^2k}{(2\pi)^2}\dfrac{e^{i\vec{k}\cdot\vec{r}}}{\eta_{N}(k^2+k/\ell_{\eta})}
%	\simeq\dfrac{1}{2\pi\eta_{N}}\log\left(\dfrac{\ell_{\eta}}{r}\right)\;.
%\end{align}

The scalar order parameter for a $\pm1/2$ disclination is roughly \mcm{constant} outside the defect core \mcm{of size $\xi_Q$, yielding} $|\bm\nabla\cdot\bm\sigma_a|\sim|\alpha|/r$ \mcm{for $r\geq\xi_Q$~ \cite{Pismen13}. Focusing} on the $+1/2$ disclinations, which are motile by virtue of self-induced active backflows\mcm{~\cite{Narayan07,Giomi13}, the divergence of the active stress $\bm\nabla\cdot\bm\sigma_a$ for a single $+1/2$ disclination} has only one non-vanishing component aligned along the axis of the defect (which we freely take to be the $x$-axis). The velocity at the center of the defect core, assumed to be passively advected by the flow, is then directed along this axis and has a magnitude given by \mcm{$u_0=\int'_{\mathbf{r}}G(r)(\bm\nabla\cdot\bm\sigma_a)_x$, where the prime indicates that the integral must be cutoff at small scales by $\xi_Q$} ($\xi_Q\sim10~\mu$m from the experimental micrographs), \mcm{below which the hydrodynamic model ceases to be appropriate, and at a long-wavelength cutoff $\ell$ controlling the screening of the $2d$ hydrodynamic flows through the coupling to the oil/water subphases. This gives
\begin{gather}
	u_0\simeq\dfrac{|\alpha|}{\eta_{N}/\ell}\mathcal{F}_>\left(\dfrac{\eta_o}{\eta_{N}/\ell},\dfrac{\ell}{\xi_Q}\right)\sim\dfrac{|\alpha|}{\eta_{N}/\ell}\ln\left(\dfrac{\eta_N/\ell}{\eta_{o}}\right)\;.
%	\left\{2+\log\left(\dfrac{\ell_{\eta}}{\ell_{\Gamma}}\right)-\dfrac{3\xi_Q}{4\ell_{\Gamma}}\left[1+\dfrac{2}{3}\log\left(\dfrac{\ell_{\eta}}{\xi_Q}\right)\right]\right\}\;,
	\label{u0}
\end{gather}
The exact form of $\mathcal{F}_>$ is given in the SI. The second approximate equality in Eq.~(\ref{u0}) holds for $\eta_{\textrm{eff}}\gg\eta_o$ and $\xi_Q/\ell\ll1$, with $\eta_{\textrm{eff}}=\eta_N/\ell$ a three-dimensional viscosity. The logarithmic dependence of $u_0$ on $\eta_o$  is robust in these limits and in agreement with the experiments (Fig.~\ref{fig_velocity}). The fit to the data, performed by means of the exact form for $u_0$ given in the SI, provides a value for $\eta_{\textrm{eff}}=13(\pm 5)~{\rm Pa~s}$ that depends very weakly on $\xi_Q/\ell$  for $\xi_Q/\ell<0.5$ (see SI).}
Since $\alpha$ is an overall scale for the defect velocity, this analysis gives an essentially parameter free estimate for $\eta_{\textrm{eff}}$. \mcm{On the other hand, the value of $\eta_N$ depends on $\ell$. A natural choice for $\ell$ is the thickness of the oil subphase ($d\sim 1$~mm) as described in the SI. This gives $\eta_N\sim13\times10^{-3}$~Pa~s~m. Alternately, one could argue that our single-defect calculation should be cutoff at the scale of the mean defect separation, which in turn depends on $\eta_o$ (see Fig.~\ref{fig_defect_density}a), albeit changing only by a factor of two ($50-100~\mu$m) over five decades of oil viscosity. Choosing $\ell\sim n^{-1/2}$, we obtain $\eta_N\sim 6.5-13\times10^{-4}$~Pa~s~m over the range of oil viscosities considered.}
%Performing the integral, we find
%\begin{gather}
%	u_0\sim\dfrac{\alpha\ell_{\Gamma}}{\eta_{N}}
%	\left\{2+\log\left(\dfrac{\ell_{\eta}}{\ell_{\Gamma}}\right)-\dfrac{3\xi_Q}{4\ell_{\Gamma}}\left[1+\dfrac{2}{3}\log\left(\dfrac{\ell_{\eta}}{\xi_Q}\right)\right]\right\}\;,
%	\label{u0}
%\end{gather}
%where $\ell_{\Gamma}=\sqrt{\eta_Nd/\eta_{w}}$ is the frictional screening length beyond which the bulk fluid acts effectively just as friction ($d$ is the thickness of the subphase) and $\xi_Q$ is
%estimated to be about $10~\mu m$ from the experimental micrographs.
% Eq.~(\ref{u0}) displays a strong logarithmic dependence of the defect velocity on $\eta_o$, in qualitative agreement with the experiments ( Fig \ref {fig_velocity}).

%We can use Eq.~(\ref{u0}) to fit the data and extract the a priori unknown active nematic viscosity $\eta_{N}$. An improved and robust fit can be obtained, using the parameter estimates obtained using the approximate expression given in Eq.~(\ref{u0}), as initial guesses to fit the data with the exact expression (see SI text and SI Fig. 1) for the defect velocity, which encodes the complete dependence on $\eta_o$. This results in an estimated shear viscosity $\eta_{N}=1.55(\pm0.02)~{\rm Pa~s~m}$ for the active MT nematic. This value can be put into perspective with the equivalent bulk viscosity of 1 nm-thick condensed phospholipid monolayers, of order $10^5$ Pa s \cite{Ding02}. In our case, we estimate a value of order $10^6$ Pa s for an active nematic layer of thickness 1 $\mu$m.

Importantly, \mcm{our fit yields $\eta_{\mathrm{eff}}/\eta_o>1$ at all but the largest oil viscosity ($\eta_o\sim10^2~{\rm Pa~s}$), where $u_0$ begins to saturate. We stress that} it is \emph{only} in this limit that the defect velocity has a logarithmic dependence on $\eta_o$. This physically corresponds to the case when the flow is dominated by the properties of the $2d$ active nematic layer and the bulk fluid only comes as a logarithmic correction to the length scale in the defect velocity. If this ratio were of order unity or smaller, as occurs at the highest value of oil viscosities, then the \emph{qualitative} dependence of the defect velocity on $\eta_{o}$ would change, giving $u_0\sim1/\eta_{o}$ instead of the strongly persistent logarithm.

In summary, we have probed the rheology of an active nematic at an oil/water interface using a set-up that allows us to vary the viscosity of the oil by five orders of magnitude. By combining experiments with a hydrodynamic model we show that measurements of the defect velocity can be used to \mcm{estimate} for the first time the shear viscosity of the active nematic.
%The value we obtain ($\eta_N\sim 1~{\rm Pa~s~m}$) is consistent with preliminary results from yet unpublished rheological measurements \cite{DogicBlair}.

%Finally, we can understand the evolution on the morphology of the active nematic with increasing oil viscosity displayed in Fig. 1 by assuming that for large values of viscosity the oil layer effectively begins to act like a frictional substrate (although viscoelastic properties will also become important). In this limit  active stresses are balanced by frictional damping and the Stokes equation for the nematic layer takes the simple form $\Gamma\mathbf{v}=\alpha\bm\nabla\cdot\mathbf{Q}$, with $\Gamma\sim\eta_o$ the friction. By eliminating the flow in the equation for the nematic order parameter, $\partial_t\mathbf{Q}=\lambda[\bm\nabla\mathbf{v}]^{ST}-a\mathbf{Q}+(K/\gamma)\nabla^2\mathbf{Q}+...$, where the superscript $ST$ denotes the symmetric traceless  part of a tensor, $\lambda$ is the flow alignment parameter, $a>0$, and $\gamma$ the rotational friction, one finds that activity yields a renormalized stiffness $K_e=K-|\alpha|\lambda\gamma/\Gamma$. The regime of turbulent form and defect proliferation occurs when $K_e<0$. In this regime the effective core size is given by $\xi_Q^e\sim \sqrt{ (|\alpha|\lambda\gamma/\Gamma-K)/a}$ and increases

\begin{acknowledgments}
We acknowledge Brandeis MRSEC Biosynthesis facility supported by NSF MRSEC DMR-1420382, and Z. Dogic and S. DeCamp (Brandeis University) for their suport in the preparation of the active gel. We thank B. Hishamunda (Brandeis University), and M. Pons and A. LeRoux (Universitat de Barcelona) for their assistance in the expression of motor proteins. We also thank J. Ort\'{\i}n (Universitat de Barcelona) and R. Casas and G. Valiente (Bluestar Silicones) for providing the silicone oil samples.
MCM was supported by the National Science Foundation through awards DMR-1305184 and DGE-1068780. SS and MCM acknowledge support from the Syracuse Soft Matter Program. P.G. acknowledges funding from Generalitat de Catalunya through a FI-DGR PhD Fellowship. Experiments where funded by MINECO Project FIS 2013-41144P.
\end{acknowledgments}

%\begin{figure}
%  % Requires \usepackage{graphicx}
%  \includegraphics[width=8.5cm]{fig_ATP_influence.pdf}\\
%  \caption{{\bf Scaling of defect density with the activity}. ({\bf a}) defect speed and ({\bf a}) defect number density in experiments performed with three different oil viscosities, 0.5 ($\square$), 5 ($\triangle$), and 300 ($\bigcirc$) Pa s.  ({\bf c}) Collapse of all data into a single master curve by using the scaling arguments discussed in the text. The straight line through the data in ({\bf c}) is a guide for the eye.}
%  \label{fig_ATP_influence}
%\end{figure}

%\bibliography{Guillamat_etal}

\end{document}